# Structure responsible for the superconducting state in $La_3Ni_2O_7$ at high pressure and low temperature conditions

Luhong Wang[1*], Yan Li[2], Sheng-Yi Xie[3], Fuyang Liu[4], Hualei Sun[5], Chaoxin Huang[6], Yang Gao[4], Takeshi Nakagawa[4], Boyang Fu[7], Bo Dong[8], Zhenhui Cao[2], Runze Yu[4], Saori I. Kawaguchi[9], Hirokazu Kadobayashi[9], Meng Wang[6*], Changqing Jin[10], Ho-kwang Mao[4, 1], Haozhe Liu[4*]

*[1]Shanghai Advanced Research in Physical Sciences, Shanghai 201203, China*
*[2]State Key Lab of Superhard Materials, College of Physics, Jilin University, Changchun 130012, China*
*[3]School of Physics and Electronics, Hunan University, Changsha 410082, China*
*[4]Center for High Pressure Science & Technology Advanced Research, Beijing 100094, China*
*[5]School of Sciences, Sun Yat-Sen University, Guangzhou, Guangdong 510275, China*
*[6]Center for Neutron Science and Technology, Guangdong Provincial Key Laboratory of Magnetoelectric Physics and Devices, School of Physics, Sun Yat-Sen University, Guangzhou, Guangdong 510275, China*
*[7]School of Materials and Energy, University of Electronic Science and Technology of China, Chengdu, Sichuan 611731, China*
*[8]Harbin Institute of Technology, Harbin 150001, China*
*[9]Japan Synchrotron Radiation Research Institute, SPring-8, Sayo-gun Hyogo 679-5198, Japan*
*[10]Beijing National Lab for Condensed Matter Physics, Institute of Physics, Chinese Academy of Sciences, Beijing 100190, China*

Corresponding authors' Emails: lisaliu@sharps.ac.cn; wangmeng5@mail.sysu.edu.cn; haozhe.liu@hpstar.ac.cn

**Abstract**: Very recently, a new superconductor with $T_c$ = 80 K was reported in nickelate ($La_3Ni_2O_7$) at around 15 - 40 GPa conditions (*Nature*, 621, 493, 2023) [1], which is the second type of unconventional superconductor, beside the cuprates, with $T_c$ above liquid nitrogen temperature. However, the phase diagram plotted in this report was mostly based on the transport measurement at low temperature and high pressure conditions, and the assumed corresponding X-ray diffraction (XRD) results was carried out at room temperature. This encouraged us to carry out *in situ* high pressure and low temperature synchrotron XRD experiments to determine which phase is responsible for the high $T_c$ state. In addition to the phase transition from orthorhombic *Amam* structure to orthorhombic *Fmmm* structure, a tetragonal phase with space group of *I*4/*mmm* was discovered when the sample was compressed to 19 GPa at 40 K where the superconductivity takes palce in $La_3Ni_2O_7$. The calculations based on this tetragonal structure reveal that the electronic states approached to the Fermi energy were mainly dominated by the $e_g$ orbitals ($3d_{z2}$ and $3d_{x2-y2}$) of Ni atoms, which are located in the oxygen octahedral crystal field. The correlation between $T_c$ and this structural evolution, especially Ni-O octahedra regularity and the in-plane Ni-O-Ni bonding angles, are analyzed. This work sheds new lights to identify what is the most likely phase responsible for superconductivity in the double layered nickelate.

## Introduction

When the result of electrical properties in nickelate $La_3Ni_2O_{7+\delta}$ under high pressure conditions up to 18.5 GPa was reported in the 21st AIRAPT conference at Cantania, Italy in Sep. 2007 [2], not much attention was attracted due to that result indicated pressure induced enhancement of the insulating phase. Drama happened after 16 years, the signature of superconductivity near 80 K in this exactly nickelate system under high pressure was reported online in July, 2023 [1], in which the high quality single crystal samples was synthesized after many year efforts [3], and the electronic occupancy of $Ni^{2.5}$ in this Ruddlesden–Popper double-layered perovskite nickelate $La_3Ni_2O_7$ might some extent mimic the $Cu^{2+}$ of hole-doped bilayer high $T_c$ cuprates, indicating that the nickel-oxide system enables the study of high $T_c$ superconductors and helps in understanding its unconventional high $T_c$ superconductivity mechanism. However, the phase diagram in this sample was plotted in the recent *Nature* paper mainly based on the transport measurement at low temperature and high pressure conditions. The assumed corresponding high pressure XRD experiments was carried out at room temperature, which was not in the P-T domains of this sample's superconducting state. The follow-up bilayer tow-orbital model mechanism investigations were based on this room temperature high pressure structure [4, 5] which might be not fastidiously proper if calculations and analysis were based on the crystalline structure at room temperature. Geisler *et al* calculated the whole group of $A_3Ni_2O_7$ (A = La – Lu, Y, Sc) as function of pressure up to 150 GPa, and proposed smaller A-site cation compound $Tb_3Ni_2O_7$ in lower symmetry *Cmc*21 structure as candidate for superconductivity at ambient pressure [6]. In contrast, structural routes to stabilize superconducting $La_3Ni_2O_7$ at ambient pressure was calculated using DFT structural relaxation based on the room temperature high pressure *Fmmm* structure model, and then proposed the routes *via* increasing the size of A-site cation [7]. Theoretical predictions seem hedge off here and might be pointing to a wrong direction and misleading many eager experimentalists in this field. Therefore, the *in situ* synchrotron X-ray diffraction experiments for this nickelate compound at low temperature region, *i. e*. at below 80 K region and pressure above 15 GPa, were carried out to check the phase transition sequence and the phase boundaries, and to determine which phase is responsible for the record high $T_c$ in this nickelate compound.

## Experimental and calculation details

The *in situ* high pressure synchrotron X-ray diffraction experiments were carried out at BL10XU, SPring-8, Japan, with the wavelength of 0.4138 Angstrom. The powder sample, which was grounded from pre-checked high quality single crystal $La_3Ni_2O_7$ sample, was loaded into sample chamber of the BeCu base diamond anvil cell (DAC) with anvil culet size of 300 micron. The details of sample syntheses were reported previously [1, 3]. The T301 stainless steel was used as gasket and silicone oil was used as pressure medium. A tiny Au foil was loaded nearby sample chunk and used as internal pressure marker [8-10]. The DAC was mounted into the online cryostat system and fully vacuumed before colling down. Helium membrane gas control system was used to control pressure remotely. Beam size was focused down to about 10 micron at sample location. Typical XRD exposure time was 1 s. Diffraction images were integrated using DIOPTAS [11], and the structural analysis and refinements was performed through GASA-II package [12].

The structural relaxations and total-energy calculations were adopted with density functional theory (DFT), which is implanted in the Vienna *ab initio* simulation package (VASP) [13]. The projector augment wave (PAW) psedopotential [14] with the energy cutoff up to 600 eV was choosen to descripe the electron-ion interation. The Perdew–Burke–Ernzerhof (PBE) functional [15] with the generalized gradient approximation (GGA) was used to calculate the exchange correlation interaction between electrons. An effective Hubbard U with 4 eV was considered in the electronic structure calculation, same as the previous related calculations set-up [1].

## Results and discussions

The sample in DAC was compressed to about 1.2 GPa at room temperature, and then performed the decreasing temperature process while collected the XRD patterns. The typical XRD patterns during this cooling process were shown in Fig. 1 (a). When temperature reached 40 K and maintained at this temperature for over 40 minutes, the gas membrane control compression process started increasing pressure slowly and the typical XRD patterns during this compression process were demonstrated in Fig. 1. (b).

A phase transition was observed when cooling down between 120 K and 104 K under pressure conditions, from starting orthorhombic *Amam* structure (space group #63, Z = 4) to orthorhombic *Fmmm* structure (space group #69, Z = 4). The typical Rietveld refinement based on this *Fmmm* structure for the XRD pattern at 15.7 GPa and 40 K was plotted in Fig. 2 (a).

When temperature was maintained at 40 K, a tetragonal phase with space group of *I*4/*mmm* (space group #139, Z =2) was discovered when compression around 19 GPa. We also performed DFT calculations and confirm the *I*4/*mmm* structure stable at pressure above 20 GPa, and the calculated optimized atomic positions in *I*4/*mmm* structure were used as the starting values in the XRD refinement for this tetragonal phase. The typical Rietveld refinement based on this tetragonal structure for the XRD pattern at 19.6 GPa and 40 K was plotted in Fig. 2 (b).

This nickelate $La_3Ni_2O_7$ belongs to so called Ruddelson–Popper (RP) series (when n = 2) $A_{n+1}B_nC_{3n+1}$ compounds, where A is usually a rare-earth, alkaline-earth or alkali ion, B can be a 3*d* or 4*d* transition metal, and C could be fluorine or oxygen. These RP series compounds are layered perovskites separated by rock-salt structure layer, as shown in Fig. 3 for n = 2 case, therefore the chemical formula could be written as $(ABC_3)_n(AC)$. The researches on these crystals family was initiated in 1950's and the very early structure founded in compound $Sr_3Ti_2O_7$ was indeed the *I*4/*mmm* structure [16], the exact structure is corresponding at the superconducting zone in nickelate $La_3Ni_2O_7$. This tetragonal structure is considered as the high symmetry parent structure in RP compounds and it has many low symmetry subgroup structures with the constraint of rigid $BC_6$ octahedra using Landau expansion [17]. At ambient conditions, the *Fmmm* structure normally is the popular type in RP compounds [18], and the nickelate $La_3Ni_2O_7$ was originally synthesized and assigned as *Fmmm* structure based on XRD data [19]. However, in 1999, high resolution neutron diffraction experiment improved the understanding on this $La_3Ni_2O_7$ structure. Due to some tiny diffraction peaks could not be explained based on *Fmmm* structure model and have solved its structure using *Fmmm* structure's subgroup: *Amam* structure [20]. *Amam* structure setting was chosen, instead of the standard *Cmcm* structure, to let the long axis as *c* direction, and make it is relatively easy for comparing with its potential higher symmetry parent structure, although the *Cmcm* setting was used in some previous literatures for $La_3Ni_2O_7$ [3].

The single crystal XRD experiments at room temperature also were performed using lab base silver source Bruker *D*8 Venture diffractometer at HPSTAR. Although the multiple twins and serve texture development easily happened when increasing pressure during multiple runs, the preliminary symmetric analysis indeed favor the *I*4/*mmm* structure model at pressure above 15 GPa at room temperature. Therefore the updated phase diagram is plotted based on these *in situ* XRD results under high pressure at 40 K and room temperature conditions, as shown in Fig. 4, in which the solid blue lines indicate the phase boundaries, and the extended blue dash lines as phase boundaries and the possible triple point were guide for the eye. This not only provides the updated phase diagram for this nickelate at low T and high P domains, but also unveils these record-setting superconductivity and the correct responsible high-pressure phase. This will raise new possibility of finding more materials that exhibit pressure-driven superconductivity with similar or much higher $T_c$ values than previously believed achievable in similar systems.

One follow-up question is how to achieve the tetragonal structure in the nickelates *via* replacement or doping other elements at A site, to make the potential high $T_c$ materials in this nickelate family become available at ambient conditions. From the point of view of the A:B atomic size ratio in this RP compounds, for example, the very first report tetragonal phase for $Sr_3Ti_2O_7$ [16], the atomic size ratio Sr:Ti = 200:140 = 1.429. For this case, La:Ni = 195:135 = 1.444, seems that smaller size atom at A site might enable to stabilize the *I*4/*mmm* structure at ambient condition. One well studied system, $La_{2-2x}Sr_{1+2x}Mn_2O_7$, with same RP structure (n = 2) and the tetragonal and orthorhombic phase could be tuned by the atomic ratio at A site, and more smaller type atoms at A site could stabilize the *I*4/*mmm* structure [21]. The tetragonal $LaSr_2Mn_2O_7$ was observed to maintain its *I*4/*mmm* structure for up to 35 GPa at room temperature [22].

The using of smaller atoms by doping A site in RP compounds might introduce chemical pre-compression effect and make the tetragonal phase become stable at ambient conditions. However, the simulated replacement test from all rare-earth elements at A site in $R_3Ni_2O_7$ provided the opposite results: The transition pressure from orthorhombic structure to tetragonal structure become significant higher in the cases with smaller atoms at A site [6]. The reversed suggestion, to replace A site with bigger atom was proposed [7], although in these calculations they were targeting room temperature structure model of *Fmmm* structure.

More calculations and experiments need to be performed to explore this effect of replacements at A site in this RP compounds.

The lattice parameters and unit cell volume as function of pressure were displayed in Fig. 5. The Z = 2 in *I*4/*mmm* structure and the double unit cell volume were plotted to reach Z = 4 for comparison with those in *Fmmm* structure. The clear volume discontinuity could be observed with the volume drop at about 1.2 % at around 19 GPa at 40 K.

The modifications of Ni-O octahedra upon compression in these RP compounds were potentially linked with the $T_c$ trend. From Fig. 6, in which the JT distortion $\sigma_{JT} = [1/6\Sigma((Ni\text{-}O)_i - \langle Ni\text{-}O \rangle)^2]^{1/2}$, it shows the further compression in the superconducting zone make the Ni-O octahedra's distortion reduced and improve regularity of the Ni-O octahedra. While the $T_c$ measurement had dropping trend with increasing pressure [1], the elongated shape of the Ni-O octahedra, especially the longer Ni-O2 bond (as shown in Fig. 3) pushes O2 atoms more interact with the A site atom at rock-salt type layer, which might have positive impact on the enhancement of the superconductivity. These nickelate samples were synthesized under about 10-19 bar oxygen gas pressure [1, 3], which potentially introduced extra oxygen to the sample structure, contrary to oxygen vacancy easily formation in other La-Ni-O systems synthesized without oxygen gas pressure. Most likely, similar to the n = 1 case in $La_2NiO_4$ system [23], extra interstitial oxygen might already get into the rock-salt type layer, which together with holes naturally play important role in the transport properties. The effect of extra oxygen needs further theoretical and experimental investigations to reveal the potential enhance mechanism on superconductivity.

It is noticed that another type of nickelates, the superconducting infinite-layer nickelates, exhibit the angle of $180^0$ at in-plane Ni-O-Ni bonding angles [24], and the correlation between $T_c$ and these angles was proposed since these angles directly impact the basal Ni $3d_{x^2-y^2}$ and O $2P_{x,\,y}$ hybridization and then the electronic structure at the Fermi energy. Experimentally, at 19.6 GPa, this in-plane Ni-O-Ni bonding angle is $178.380^0$ and slightly increased to $178.397^0$ when pressure compressed to 24.6 GPa at 40 K in $La_3Ni_2O_7$. This seems not really in favor of the proposed positively correlation for the $T_c$ trend upon compression in this RP compound. However, attention need be paid to avoid over-explanation on the crystalline structural detailed information from the XRD refinement results, since the diffraction quality under two extreme conditions, high pressure and low temperature simultaneously, normally became worse compared to those obtained at ambient conditions.

The fundamental building block of the 3 structures (space group *Amam*, *Fmmm*, and *I4/mmm*) for $La_3Ni_2O_7$ in this P-T range are basically similar. However, the detailed electronic structures demonstrate different features. The DFT calculations show that the *Fmmm* structure will spontaneously transform to the *I4/mmm* structure at 20 GPa without any barrier during the structural relaxation. The total energy varied with the volume of *I4/mmm* structure at 20 GPa was plotted in Figure 7 (a) and the projected band structure was illustructed in Figure 7 (b). The electronic states approached to the Fermi energy were mainly dominated by the $e_g$ orbitals ($3d_{z2}$ and $3d_{x2\text{-}y2}$) of Ni atoms, which are located in the oxygen octahedral crystal field. The Ni atoms' low energy $t_{2g}$ orbitals ($3d_{xy}$, $3d_{yz}$ and $3d_{xz}$) in octahedral field hybrize effectively with $2p$ orbitals of oxgen and sink deeply away from the Fermi level. These results agree with previous studies in general [1, 6, 7], which indicates that the Copper pair of superconductivity is derived from the $e_g$ orbitals of Ni. However, the detailed mechanism of pairing should be further investigated.

In RP type nickelate system, when n = 3, the $T_c$ of 21 K at about 70 GPa high pressure conditions at $La_4Ni_3O_{10}$ system were reported very recently [25, 26]. These results enrich the understanding of superconductivity in RP type nickelates. However, no *in situ* structure measurements were performed in the $La_4Ni_3O_{10}$ system so far, and again it is missing important piece of information on the corresponding structure, which is crucial for solving mechanism puzzle. These updates, together with previous research on this system [27], will bring more interesting subjects on the researches to explore the universal relationship between structure evolution and the properties change in RP type La-Ni-O family at low temperature and high pressure conditions.

## Conclusions

In summary, the *in situ* high pressure and low temperature synchrotron XRD experiments were carried out and we solved the responsible structure in $La_3Ni_2O_7$ for its high $T_c$ state. At 40 K, a tetragonal *I*4/*mmm* phase was discovered when compression around 19 GPa. The correlation between $T_c$ and this structural evolution, especially Ni-O octahedra regularity and the in-plane Ni-O-Ni bonding angles, are discussed. This updated phase diagram based on XRD results, will help us better understand the mechanism in this unconventional high $T_c$ superconductor in $La_3Ni_2O_7$, and offer relatively clearer guideline to synthesize related compound systems, and potentially eventually realize the superconducting state, instead of at high pressure conditions, at ambient conditions.

**Acknowledgments:** This work was supported by the Natural Science Foundation of China (11374075, 12174454, and 11704111). The authors acknowledge financial support from Shanghai Science and Technology Committee, China (No. 22JC1410300) and Shanghai Key Laboratory Novel Extreme Condition Materials, China (No. 22dz2260800). Work at SYSU was supported by Guangdong Basic and Applied Basic Research Funds (grant no. 2021B1515120015), Guangzhou Basic and Applied Basic Research Funds (grant no. 202201011123), and Guangdong Provincial Key Laboratory of Magnetoelectric Physics and Devices (grant no. 2022B1212010008). Calculation resources were provided by the National Super-computing Center in Shenzhen (Shenzhen Cloud Computing Center).

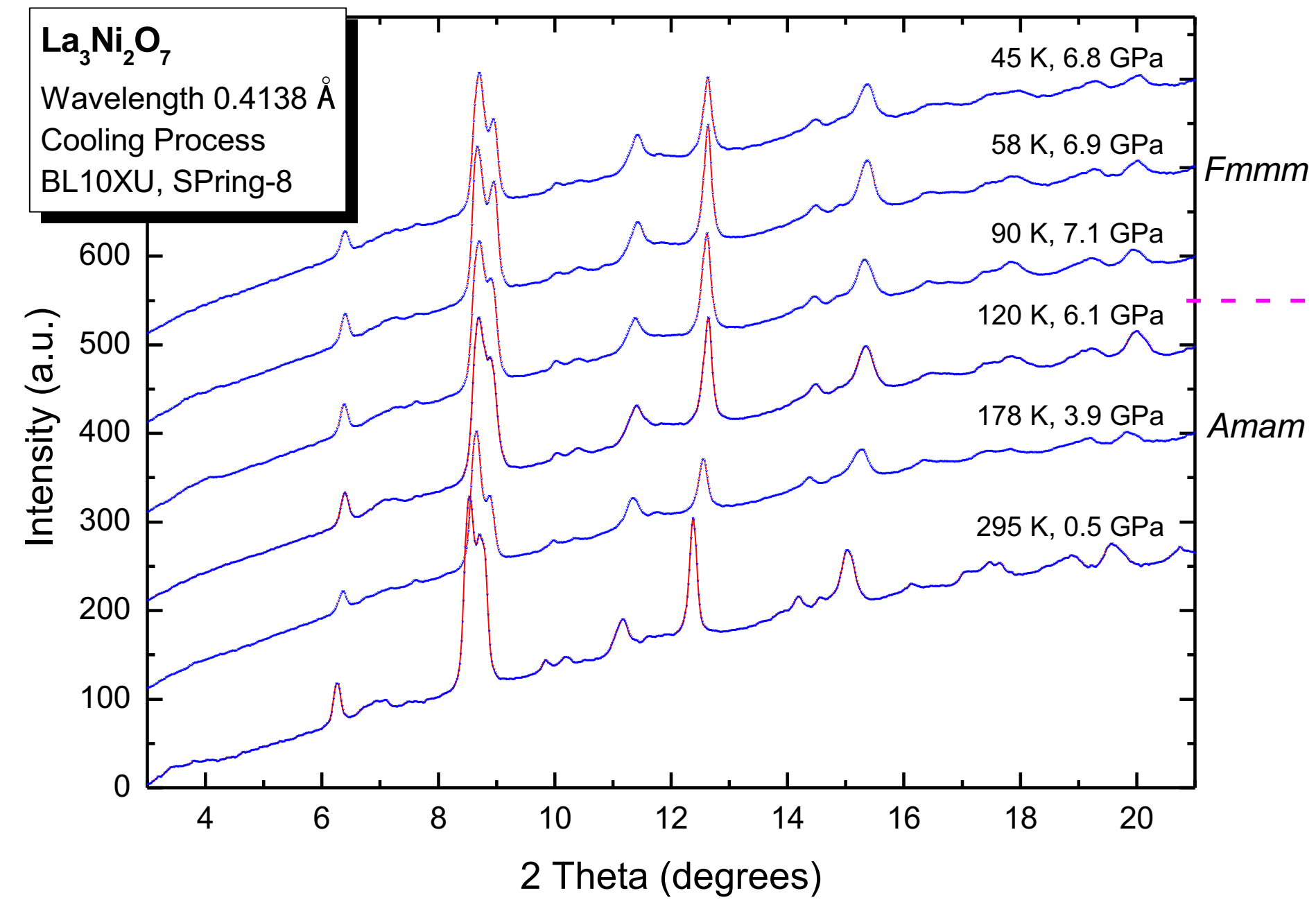

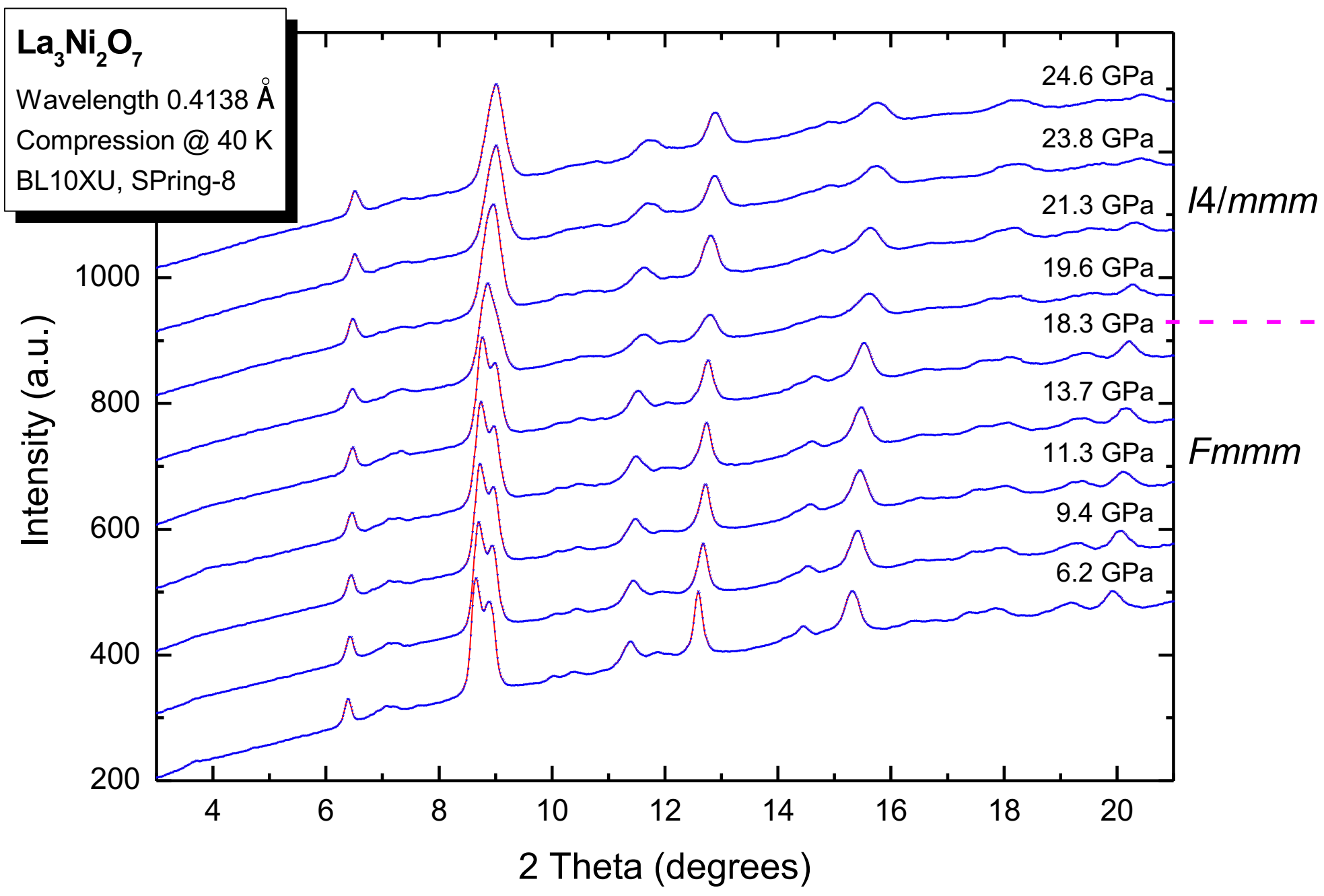

Fig. 1 Selected typical XRD patterns during (a) cooling process, and (b) compression process at 40 K.

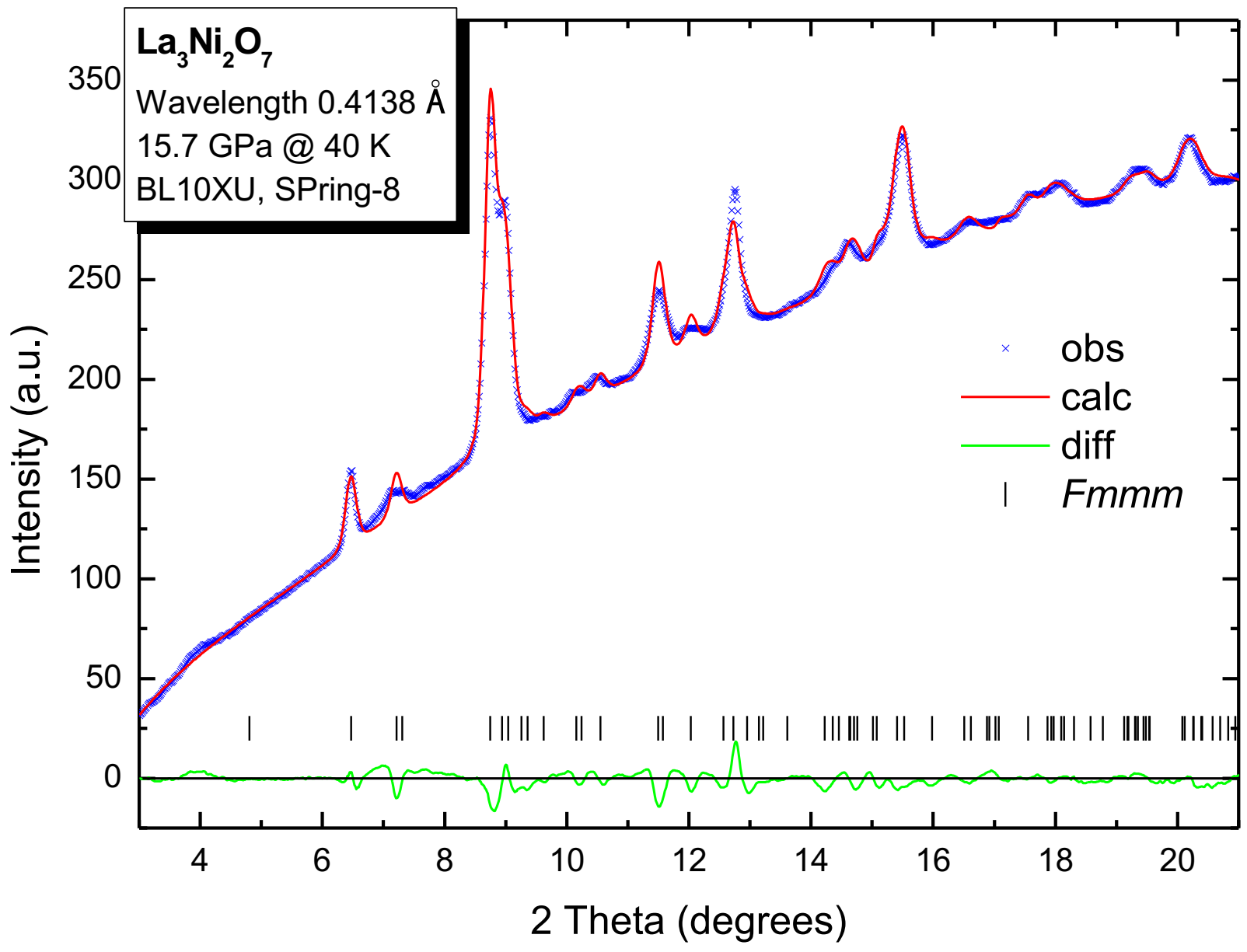

Fig. 2 (a) The typical XRD refinement result for *Fmmm* phase at 15.7 GPa and 40 K, $R_w$=1.164%.

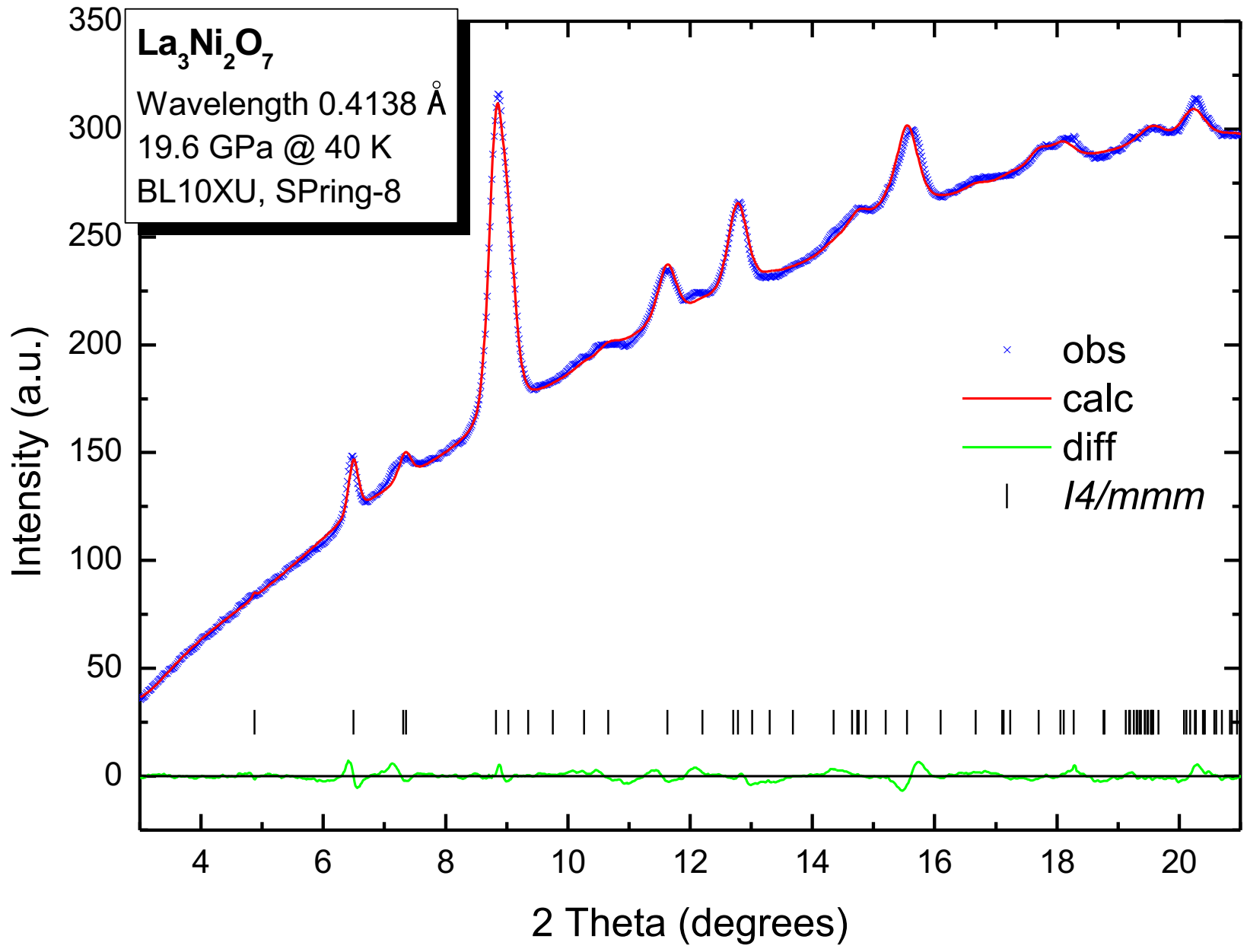

Fig. 2 (b) The typical XRD refinement result for *I*4/*mmm* phase at 19.6 GPa and 40 K, $R_w$=0.656%

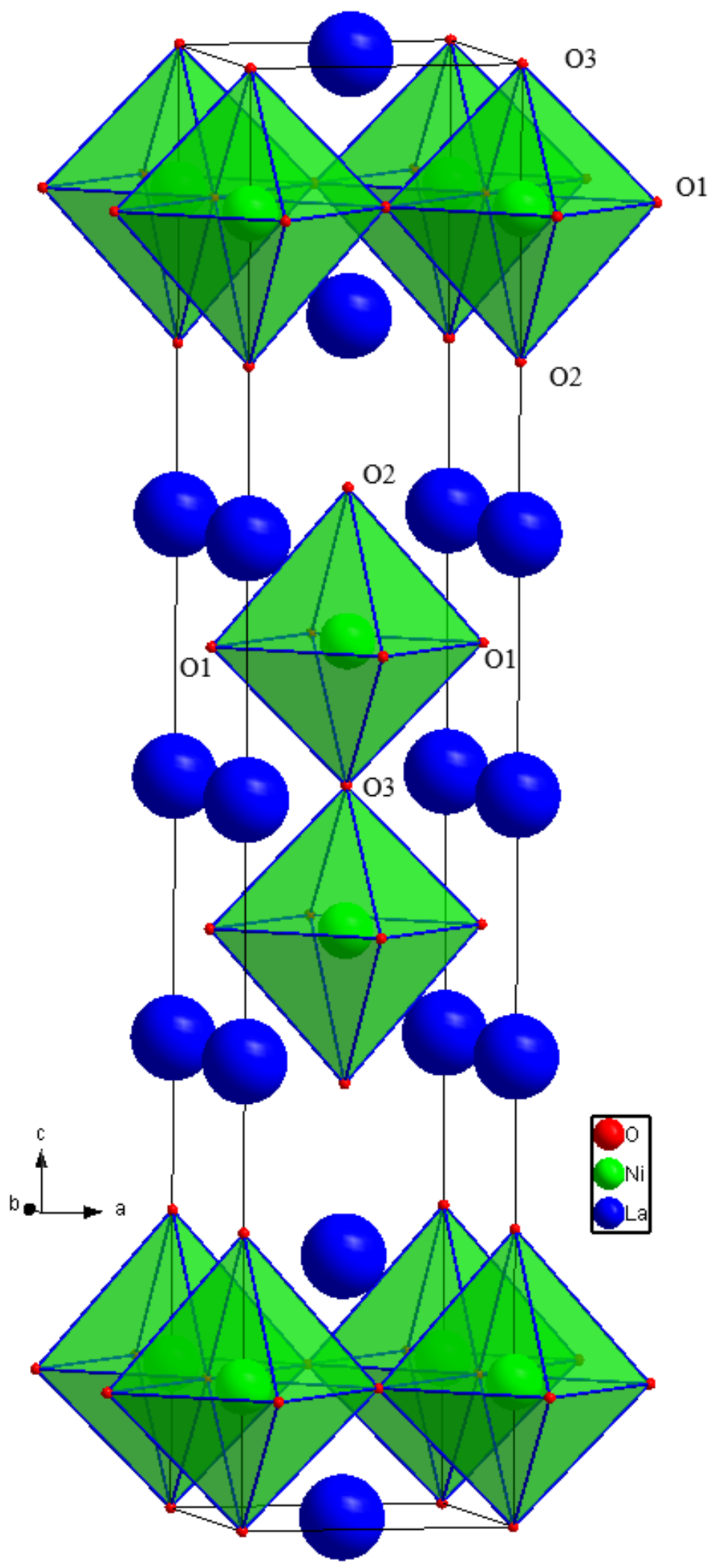

Fig. 3. Unit cell (Z=2) of *I*4/*mmm* structure for $La_3Ni_2O_7$

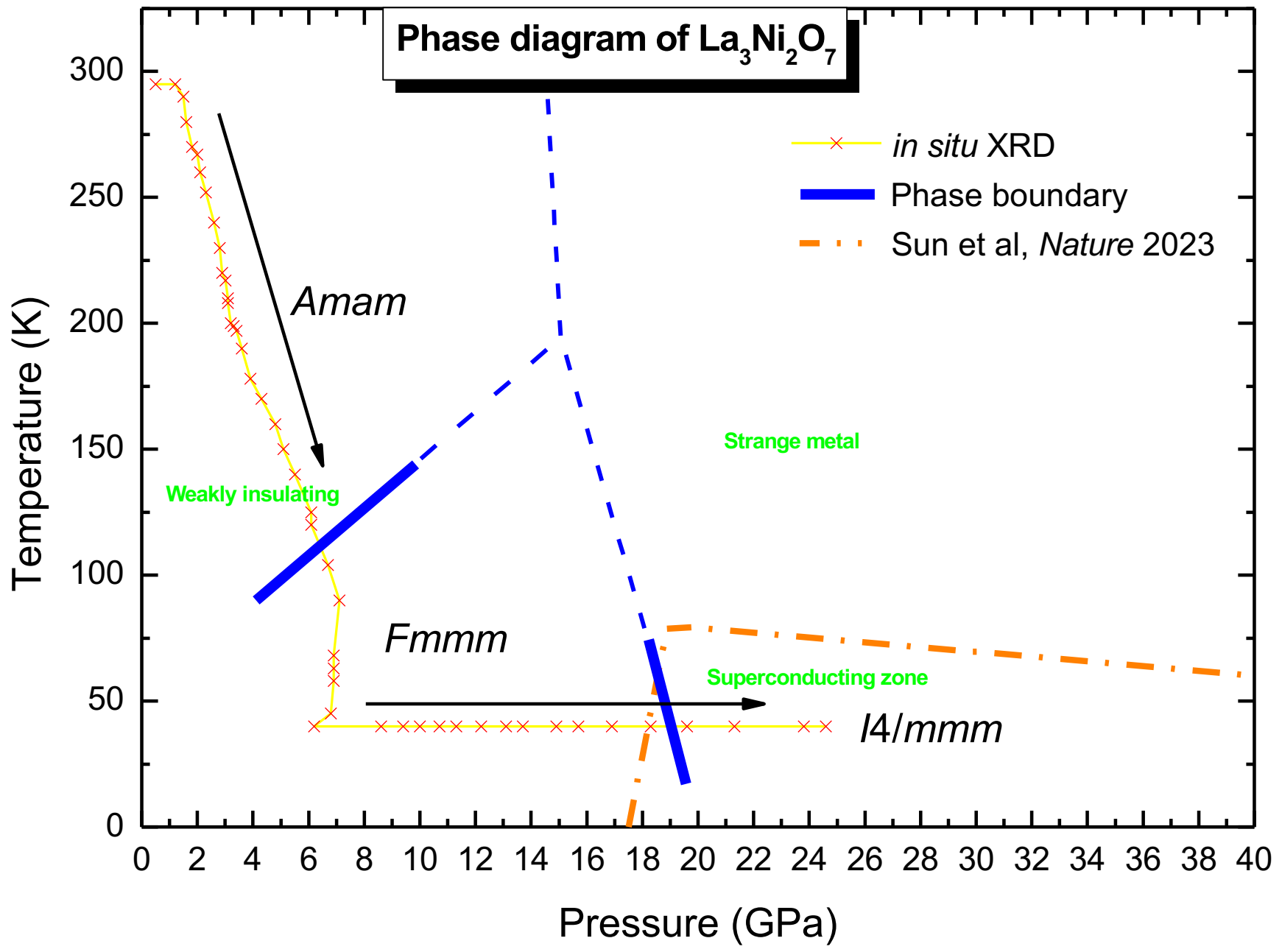

Fig. 4. Phase diagram of $La_3Ni_2O_7$ based on the *in situ* XRD results, in which the black arrows show the path of the cooling and compression process.

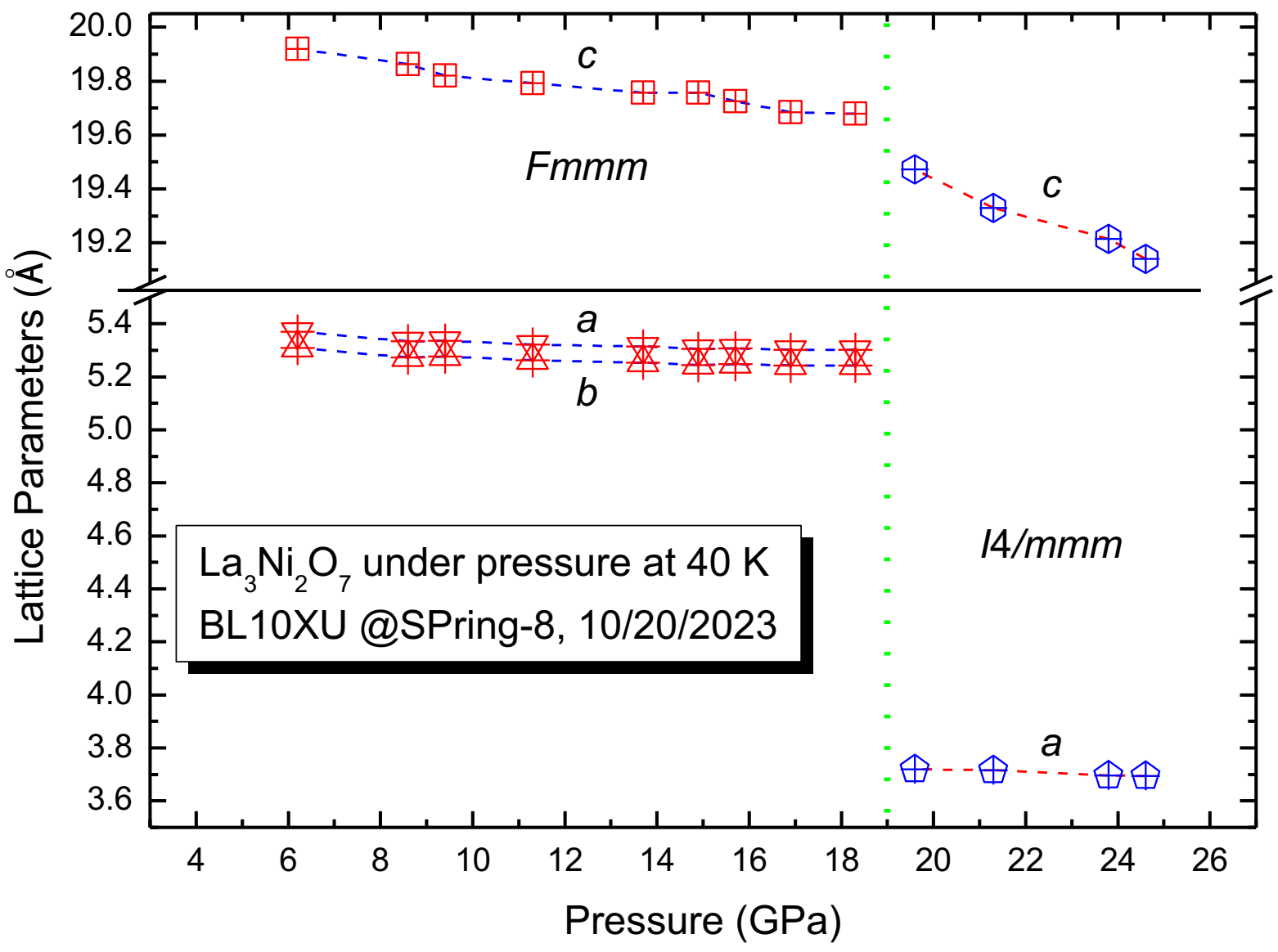

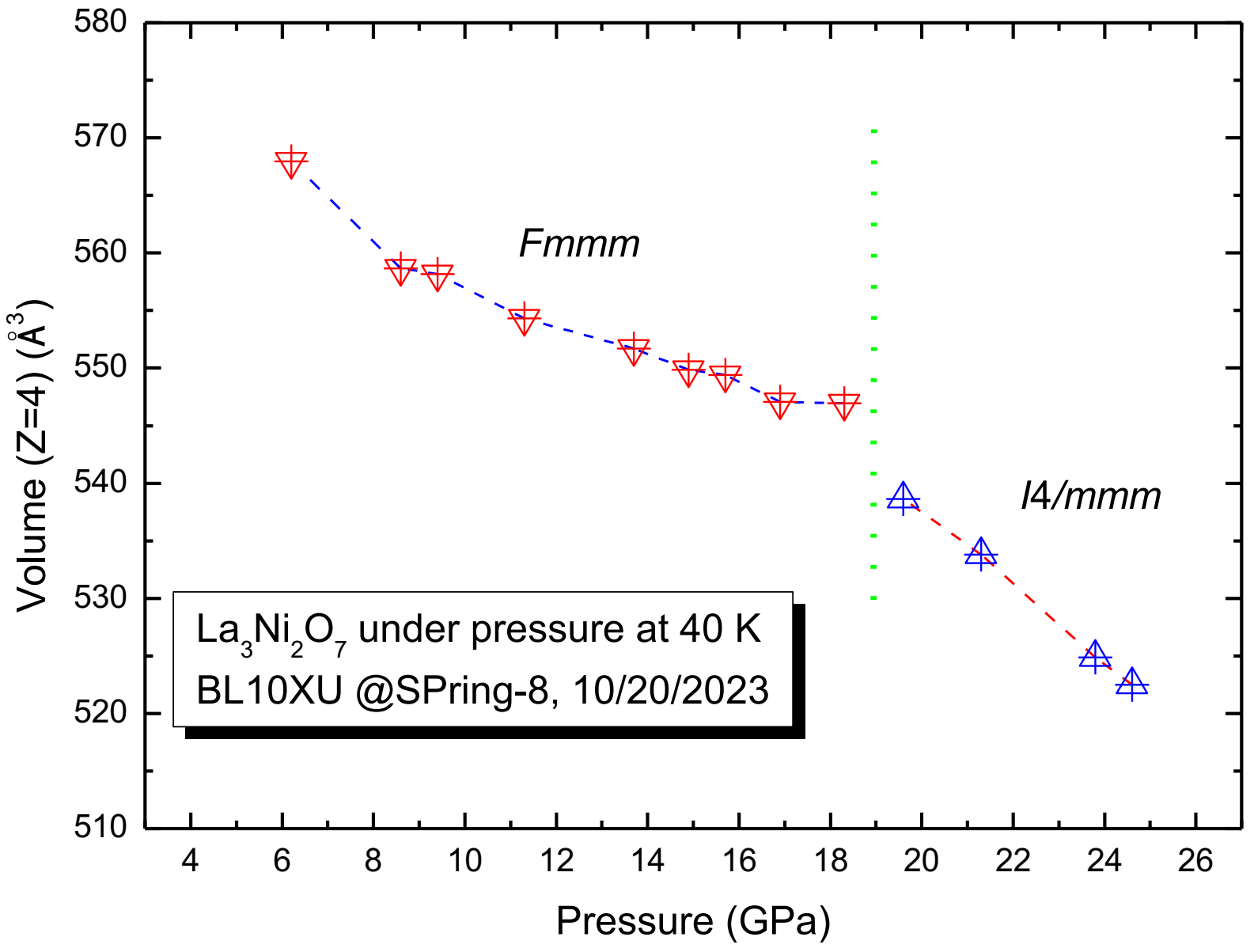

Fig. 5 (a) Lattice parameters, and (b) volume ($Z = 4$) as function of pressure at 40 K.

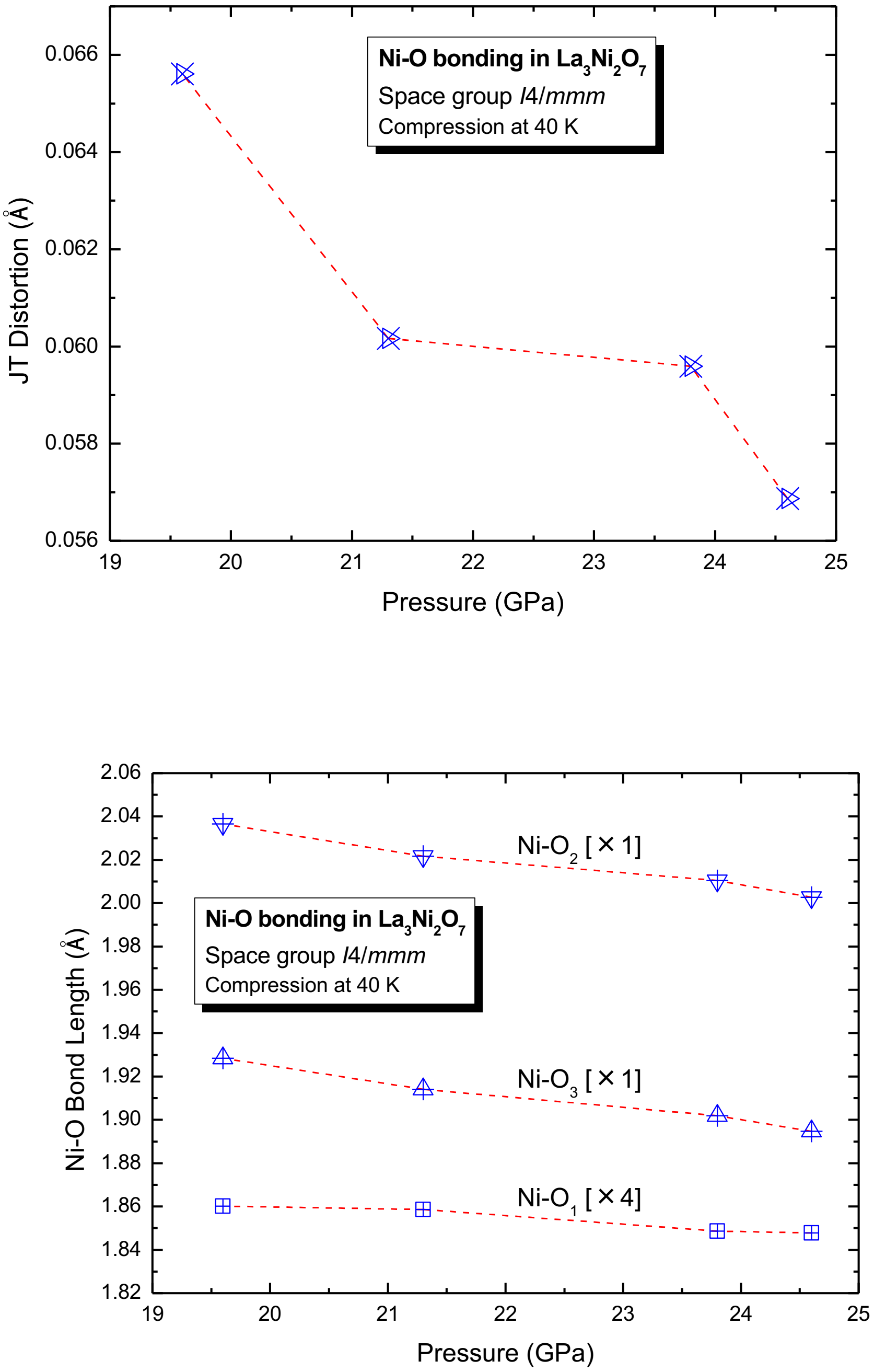

Fig. 6 (a) Regularity of Ni-O octahedra, in which the JT distortion $\sigma_{JT} = [1/6\Sigma((\text{Ni-O})_i - \langle\text{Ni-O}\rangle)^2]^{1/2}$, and (b) the Ni-O bonding length in *I*4/*mmm* structure as function of pressure at 40 K.

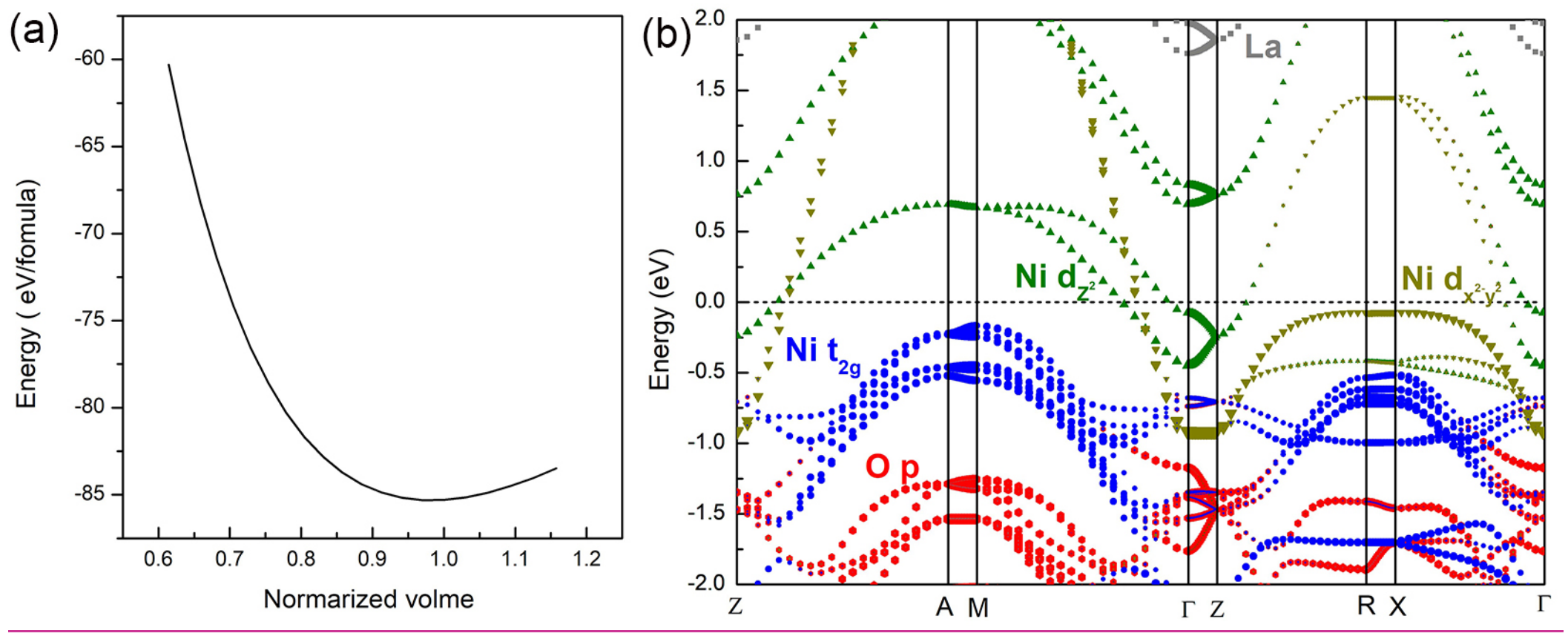

Fig. 7 (a) The variation of energy with the volume of *I*4*/mmm* structure; (b) Projected band structure of *I*4*/mmm* structure at 20 GPa.